\documentclass{llncs}

\usepackage{times}
\usepackage[pdftex]{color, graphicx}
\usepackage{subfigure}
\usepackage{algorithm}
\usepackage{algorithmic}

\title{SuperNova: Super-peers Based Architecture for Decentralized Online Social Networks}

\author{ Rajesh Sharma and Anwitaman Datta \institute{School of Computer Engineering, Nanyang Technological University,
Singapore. \\ \{raje0014,Anwitaman\}@ntu.edu.sg}}

\begin{document}
\maketitle

\begin{abstract}
Recent years have seen several earnest initiatives from both academic researchers as well as open source communities to implement and deploy decentralized online social networks (DOSNs). The primary motivations for DOSNs are privacy and autonomy from big brotherly service providers. The promise of decentralization is complete freedom for end-users from any service providers both in terms of keeping privacy about content and communication, and also from any form of censorship. However decentralization introduces many challenges. One of the principal problems is to guarantee availability of data even when the data owner is not online, so that others can access the said data even when a node is offline or down. Intuitively this can be solved by replicating the data on other users' machines. Existing DOSN proposals try to solve this problem using heuristics which are agnostic to the various kinds of heterogeneity both in terms of end user resources as well as end user behaviors in such a system. For instance, some propose replication at friends, or at some other peers based on other heuristics such as reciprocal storage among nodes with similar availability, or storage in a global DHT realized using all peers' resources. In this paper, we argue that a pragmatic design needs to explicitly allow for and leverage on system heterogeneity, and provide incentives for the resource rich participants in the system to contribute such resources. To that end we introduce SuperNova - a super-peer based DOSN architecture. Super-peers can help (i) bootstrap new peers who are yet to have/find any friends by either providing them storage space, (ii) maintaining a directory of users, so that users can find friends in the network by name or interests, (iii) help peers find other peers to store their content in case they don't have adequate friends to do so, or if their friends are already overloaded. We envision a self-organizing system, where nodes that provide substantial resources can gain reputation, and be elevated to the status of super-peers. Users may want to become super-peers out of altruism (they want DOSNs to succeed), for the sake of the reputation (e.g., being an influential member for an interest based community) as well as potentially to monetize their special roles (e.g., run advertisements). While proposing the SuperNova architecture, we envision a dynamic system driven by incentives and reputation, however, investigation of such incentives and reputation, and its effect on determining peer behaviors is a subject for our future study.  In this paper we instead investigate the efficacy of a super-peer based system at any time point (a snap-shot of the envisioned dynamic system), that is to say, we try to quantify the performance of SuperNova system given any (fixed) mix of peer population and strategies.
\end{abstract}


\section{Introduction}
\vspace{-10pt}
Online social networks (OSN) such as Facebook are extremely popular,
and continues to
grow\footnote{http://www.facebook.com/press/info.php?statistics},
and has grown from the early day use case of being just a tool to
keep in touch with friends into something much more pervasive in
both our private life and work place, including organization or
promotion of group events or orchestration of collaborative tasks.
With the increasing number of users, and user centric personal data
online, such networks however raise various privacy concerns.
Concerns include inadequate or confusing privacy settings to share
information with fellow OSN users (which can be addressed using
flexible access control policies supported with simpler user
interfaces), as well as more fundamental issues related to data
ownership and privacy and autonomy from the social networking
service providers themselves, as well as possible vulnerability of
such vast repositories of personal data to identity harvesters. A
slew of initiatives, originating both from academic research such
as PeerSoN\cite{peerson}, Vis-a-Vis \cite{Vis-a-vis}, SafeBook
\cite{safebook} as well as open-source and privacy/free speech
centric communities such as
Diaspora\footnote{http://www.joindiaspora.com/},
OpenSocial\footnote{http://www.opensocial.org/},
NoseRub\footnote{http://noserub.com/} have in the last couple of
years started looking at decentralized architectures which can
provide online social networking services without the trappings of
such services being run by centralized providers.

The path to the decentralization of online social networking itself
is however not an easy one to tread. Using end user resources in a
peer-to-peer manner does not naturally mean scalability or
robustness in terms of quality of services, on the contrary, an
internet-scale decentralized online social network run by autonomous
and individually unreliable peers is bound to have poorer
performance if compared with a well provisioned, centrally managed
dedicated infrastructure based service. Likewise, a peer-to-peer
approach does not naturally provide privacy or security, on the
contrary, due to the lack of any accountable authority, running
peer-to-peer software may be more vulnerable to malwares. Finally,
not everyone is equally passionate about privacy issues, and in the
absence of a critical mass, even if security related technical
challenges to realize a decentralized online social network (DOSN)
are overcome, the exercise may be futile. Like any other business
idea, it is however difficult to foresee whether a specific product
will appeal to a significant market base. Nonetheless, we argue,
that similar to the usage of PGP (pretty good privacy) for email
based private communication, or Tor \cite{Tor} for anonymous
communication over the internet, while the percentage of users who
do use such techniques is miniscule, the absolute number of people
who perceive the need for the technology to facilitate the same is
significant. Thus, scientifically, it is imperative to explore and
create the technology which can support decentralized online social
networks. Both the philosophical aspects, as well as very many
technical challenges, as well as opportunities of realizing DOSNs
have been extensively discussed in \cite{buchegger:case} and other
DOSN related publications \cite{peerson,safebook,Vis-a-vis,porkut},
so we don't go into further discussions on the merits and
motivations of DOSN in this paper, and instead primarily focus on a
subset of concrete technical issues that we try to address. It is
also worth pointing out at this juncture that, besides being used by
privacy conscious netizens in their daily life as an alternative to
a centralized OSNs, DOSNs may also appeal to corporate entities, who
may want to benefit from the advantages of online social networking
to enhance the productivity of their employees, and yet would be
apprehensive to put corporate data on a public OSN like Facebook (a
comparison of corporate emails with free email services may be
made). The super-peer based DOSN architecture we propose naturally
fits such corporate needs as well, besides catering to the privacy
conscious individuals.

Among other services, a centralized OSN system uses dedicated
servers to store user data (which includes user profile, other
content such as pictures and video files, as well as communication
with other users). In a decentralized setting, ensuring 24/7 data
availability is non-trivial. Replicating the
data\footnote{Replication instead of (erasure) coding based
redundancy techniques is more suitable for storing OSN data, since
size of individual data is typically small, while access and
manipulation of such data is relatively frequent. Erasure coding is
more suitable for large static data for archival and back-up
applications with relatively infrequent accesses.} over the network
is necessary to increase availability. However, the choice of
appropriate network nodes where to carry out such replication is
neither obvious, and may further be complicated by various issues
such as resource constraints and individual nodes' behaviors. Even
if some enforcement mechanism can be devised such that every
participant node cooperate together to increase the overall
availability of data in the whole system (common good),
computationally it is (NP-)hard \cite{krzicdcs10}. In a
decentralized setting, such enforcement is likely infeasible, and if
incentive mechanisms such as reciprociyt
\cite{pietrogametheory,krzicdcs10} are used, (rational) peers would
pair-up with other nodes greedily to maximize their own data
availability, such that the price of anarchy in the system is
unbounded \cite{krzicdcs10}, that is to say, highly available nodes
pair up together, making their own data highly available, while all
other nodes in the system fail to benefit from the existence of such
well provisioned nodes, and suffer from poor availability of data.

Consequently, the proposed super-peer based architecture is based on
the key observation that it should be possible to incorporate some
other forms of incentives in the system. A super-peer based
architecture benefits in that, (i) it allows the nodes acquiring
super-peer status to capitalize on the same in various manners, such
as monetize by serving advertisements to the serviced peers, or gain
recognition and influence a community; and (ii) it allows for an
alternative form of incentive for peers to behave well, in order to
gain such super-peer status, in order to in turn benefit from the
perks of being a super-peer. Furthermore, a super-peer based
architecture allows explicit mechanisms to leverage on well
provisioned nodes, which some users may want to contribute, either
because of an ulterior incentive that they would like a DOSN to be
sustainable (similar to volunteer relay nodes in Tor), or being run
by corporate entities who would like to benefit from the advantages
of social networking, while not lose control of data\footnote{In the
case of corporate run super-peers, corporate super-peers would take
care of all storage needs for their respective employees}.
Furthermore, such `superpeers' may be run either on end user
computers, or on cloud services. Finally, having such super-peers
help facilitate new node joins in the system, since new joinees
would typically not immediately know enough friends. This is a
critical functionality, which is often ignored in existing
literature \cite{peerson,soja,safebook,porkut}, which deal with data
management issues assuming that nodes have already established
(adequate) social connections.

In this paper, we make the following contributions.
\begin{itemize}
  \item We propose a new architecture for realizing decentralized online social networks (DOSNs) which is a super-peers based network of volunteer agents (hence the name, SuperNova). We have arrived at such a super-peer based design guided by our past experiences in implementing and experimenting with the PeerSoN DOSN \cite{peerson} and other decentralized social information systems \cite{soja} relying on other `flat' architectures, as well as based on insights obtained from some theoretical works \cite{krzicdcs10}. The proposed SuperNova architecture allows the system to self-organize by leveraging on heterogeneity, which may arise due to various reasons including altruistic behavior of participants as well as various incentive mechanisms.
  \item We note that the underlying incentive mechanisms will lead to a dynamic system, where peers' reputation and role in the system will be in continuous flux, and is a topic worth studying in its own right. We defer such a study for later. The focus of this paper is restricted to the algorithmic and systems design and evaluation aspects.
  \item We investigate (based on simulations) the performance of the proposed system for a static snap-shot, which results from a mix of strategy choices made by the constituent user population. Such extensive exploration (albeit of only a part) of the DOSN design and parameter spaces is in contrast with most existing DOSN literature, which have focused on the system design and security protocols or have evaluated small scale deployments in controlled environment. While such evaluations are also important, they are inadequate to determine the fundamental impact of specific design decisions and system parameters (determined, for instance by user strategies and capacities) or to discern important design concepts, and mainly manifest the performance of one specific implemented artefact under a specific controlled environment. In contrast, the evaluations in this paper help determine which mixes of user populations may lead to sustainable DOSNs, versus which mixes are not sustainable, as well as help quantify the achieved quality of service in each scenario. More specifically, the different user strategies include: choice of nodes where to store its data which can be at friends, at strangers, at super-peers or a combination of the same, as well as the choice of nodes to determine whose data they are willing to store, again including friends and/or strangers; besides nodes' storage capacity and churn, which all together determine the system environment.
\item In this paper, we propose mechanisms to judiciously utilize user resources, given a mix of user strategies and contributed resources, by detailing the various resource management tasks played by the super-peers.
\end{itemize}

Security issues, which will be essential for practical deployment of a SuperNova like system, is out of the scope of this paper. Ideas explored in Plutus \cite{plutus}, SafeBook \cite{safebook} and \cite{hungACSAC} are expected to be a starting point for the same. We further discuss related works next in Section \ref{sec:relwork}, where we elaborate both on how this paper contributes a new
architecture in the context of DOSNs, as well as explain how our
evaluation provides practical insights missing from earlier works.
We present the SuperNova approach including various concepts and
algorithms in Section \ref{sec:approach}. Section \ref{sec:eval}
evaluates the performance of such a system using synthetic and
real-life social network graphs, and for various mixes of user
strategies and capacities. We draw our conclusions in Section
\ref{sec:concl} along with an outline of several future
directions we intend to pursue.

\section{Related works}\label{sec:relwork}
\vspace{-10pt}
Work on peer-to-peer storage systems date back to the OceanStore \cite{oceanstore} initiative to achieve archival storage using end-user resources. More than a decade of P2P storage related research later, we have several hybrid or peer-to-peer storage and backup cloud like services in actual deployment, e.g., Wuala \footnote{www.wuala.com}, where storage task is centrally coordinated, or as academic prototypes such as FriendStore \cite{friendstore}, where storage is carried out at nodes' friends. While allowing for options for sharing and socializing, the original design of such systems is not social networking. At a very high level, other instances of hybrid (central coordination assisted) peer-to-peer virtual community networks include internet telephony service like Skype \cite{Skype}, P2P massively multiplayer online games and virtual worlds \cite{P2PMMOG} and social peer-to-peer file sharing systems like Tribler \cite{Tribler}. The need and challenges of realizing DOSNs were formalized recently \cite{buchegger:case}, and since then, there has been a flurry of academic initiatives to realize both complete systems, as well as to surmount individual challenges. A more exhaustive survey of existing DOSN specific research covering various aspects of DOSN designs can be found in \cite{DOSNsurvey}.

Some of the prominent (and architecturally representative)
prototypes of DOSN include PeerSoN \cite{peerson}, SoJa \cite{soja},
SafeBook \cite{safebook} and Vis-a-Vis \cite{Vis-a-vis}. The PeerSoN
design, in principle assumes a global storage service realized using
user resources (based on a DHT), but storage location and
responsibility is disentangled from social relations. SoJa is a DOSN
based social library application, which identifies a dichotomy of
storage services - namely global and `socially' local storage, where
the global storage is again realized using a user resource
contributed global DHT, and stores interesting information such as
user directory, as well as allows storage on encrypted offline
messages for users (similar to as in PeerSoN), while nodes store
data which will typically interest only friends in the local storage
realized using friend nodes' resources. Data security is the primary
focus of Safebook \cite{safebook}, where the degree of trust among
nodes is used to create a multi-ring system called Matryoshkas. In
Vis-a-Vis \cite{Vis-a-vis} each individual users host its data on a
virtual individual server (Vis), which is run on a cloud service.
Social relation agnostic storage approaches like PeerSoN do not
allow for incentive mechanisms for users to contribute resources.
Approaches leveraging on only immediate social and trust relations
(as in SoJa, FriendStore, SafeBook), or based on reciprocity
\cite{krzicdcs10} provide incentive mechanisms with limited scope,
since users benefit only within the confines of their individual
social neighborhoods, rather than at the level of a community at
large, and such limitations lead to poor level of availability
\cite{P2P-localityIssues,krzicdcs10}.

In Vis-a-Vis, each peer needs to provision for its storage needs, possibly by making monetary
payment to a storage cloud service. In SuperNova, we try to leverage
on the advantages of each of these approaches, by allowing a
multitude of storage and strategy choices - and by adapting an
architecture which allows both bilateral as well as community wide
incentive mechanisms. The super-peers in SuperNova help realize a
peer provisioned storage cloud, by both providing storage
themselves, as well as by help match nodes to meet storage needs. In
turn, the super-peers themselves may, similar to Vis-a-Vis, be run
either on commercial cloud services, or by running (relatively)
dedicated personal computers by the end users. Users can store their
data at friends, super-peers or at `stranger nodes'. For the later,
matching, coordination and accounting being carried out by
super-peers. Likewise nodes may choose to allow friends or strangers
store data on them, in order to gain reputation and influence in the
community, which can culminate into super-peer status, which in turn
allows opportunities to monetize the service. Other extrinsic
incentives or reasons for nodes to run super-peers may be
altruistic, or vested interests such as corporate entities running
super-peers to manage their employee `accounts'. Thus, what we
achieve with the SuperNova architecture is in some sense a generic
hybridization of the existing architectures, allowing the users to
participate in an ecosystem, where each individual user has the
flexibility to choose different trade-off points.

Another recent approach looks at using user resources to help scale
centralized OSNs \cite{UserAssisted-OSN} by using user resources to
distribute large files among friends, but in such a scenario, the
users' role is more that of a P2P content distribution network, and
the privacy and autonomy issues of centralized OSNs persist.

Besides the fundamental architectural and design differences with existing DOSNs, the presented work has several other differences with respect to the existing literature. We provide explicit mechanisms for new nodes to join the system, and leverage on the resources provided by super-peers and strangers, while they wait to establish or discover adequate friendship relations.

Finally, we carry out experiment based evaluation of different parameter choices to determine quantitatively the impact of different mix of user strategies on the sustainability of DOSNs, which contrasts with evaluation of existing prototypes, which mainly focus on functional correctness in controlled settings, rather than exploration of the design parameters and system environment (as we do).

\section{Approach}\label{sec:approach}
\vspace{-10pt}
In a DOSN system, to increase availability a particular node $n$'s data has to be replicated among other nodes which are selected by exploring various social links - friendship and strangers (with help from super-peers). We explore best-effort storage mechanisms to increase (rather than 24/7) availability.

In this section we first describe various entities of the system before describing the bootstrap process for a new node. We also describe different strategies that a node might select for increasing its data availability, followed by our system model.

\vspace{-12pt}
\subsection{Definitions of Different Entities}
\vspace{-7pt}
 This section describes various entities which are parts of the SuperNova architecture. A node (or user) $n$ in a DOSN has a list of friends. Some of the friends might be ready to store friend's data thus acting as storekeepers (We interchangeably use (1) node and user and (2) friends and neighbors in this paper.).

\vspace{-12pt}
\subsubsection{Profile:}
\vspace{-7pt}
Each node (or user) in DOSN is represented by a profile which can be considered as a medium of expression for the user, as well as user data - which may be public, or protected for limited access to a subset of friends, or private and inaccessible to anyone (for example, for back-up).

\vspace{-14pt}
\subsubsection{Friendlist:}
Each user's profile is linked with his friend's profile,
which user adds either based on acquaintances or on interest similarity.
The collection of all the linked profile is termed as friendlist. We assume friendship is a symmetric relationship. 

\vspace{-14pt}
\subsubsection{Storekeepers:}
Storekeepers are list of users who have agreed to keep a replication of another user's data so that when a particular node $n$ is down, then $n$'s friends can contact storekeepers to access $n$'s data. In other words, $n$'s storekeepers give visibility to $n$, when it is down. We describe the process of selection of storekeepers by node $n$ in section \ref{sec:protocol} in more detail.
Not every friend is willing to act as storekeeper because of prior load or due to personal reasons. Every storekeeper maintains a list of nodes for which they are doing a storekeeping for data synchronization. Unlike friendship, storekeeping is not a symmetric relation. For a particular node $n$, every friend in his friendlist knows the list of all the storekeepers (described next) for the profile (users). During initial period of time when a node $n$, doesn't have enough friends it requests one of the super-peers to store his data (see section \ref{sec:bootstrap4node}). Thus for new nodes super-peers act as storekeepers.

Each node $n$'s profile data is divided into three parts namely - Public (Pb), Protected (Pt) and Private (Pv). Each node $n$ has list of friends $F_{1}$, $F_{2}$, ..., $F_{f}$ with which his profile is linked. Each friend $F_{x}$ $(1 \leq x \leq f)$ has information about the list of storekeepers ($SK_{1}$, $SK_{2}$, .. ,$SK_{sk}$) which are storing node $n$'s protected data. Each storekeeper $SK_{y}$ $(1 \leq y \leq sk)$ has information about all the other storekeepers which are storing node $n$'s data. Friends can access Protected and Public data (enforced using mechanisms described in \cite{plutus} and \cite{hungACSAC}). Public data can be accessed by any node in the network whereas private data is solely for node $n$ itself.


\vspace{-14pt}
\subsubsection{Super-peers:}
Super-peers are one of the most important entities in the system. Any node may become a super-peer by providing his services to the system in general, and most importantly to new nodes. Super-peers provide storage to the new nodes who don't have enough friends in the network for some initial period. They also maintain and manage different types of services (for example maintenance of user-list) for DOSN by cooperating among themselves.

In the envisioned system, different super-peers may provide different kind of services. Some of them might be offering good storage and bandwidth resources, others might be carrying out good quality of book keeping, recommendations and match making (e.g., to help new users find suitable friends with shared interests, or find reliable storekeepers) and some a combination of these. We note that appropriate incentive and reputation based mechanisms are needed to bring accountability and allow the super-peers to be fairly rewarded for their contributions.

\vspace{-14pt}
\subsubsection{Communities:}\label{sec:community}
Groups or communities are formed based on interest. Any user can create a community. The creator of the community will act as a moderator for the community. Creator (or owner) of the community is the main responsible person for storing all the data and for settling any dispute if it arises in the community due to any posting or because of any other reason. A creator (and owner) may share/delegate his ownership with other nodes if he wishes so, e.g., for load-balancing. Depending on rules and regulations set by the moderators, community members will have different access for the community data.

The bulk data related with community will be shared by other members of the community as well. However community moderators take care of all the data
(including old and new). There are two ways to define the latest data, either w.r.t. time or storage space. For instance, in terms of time,
latest data could be of last 1 week. In terms of space it could be latest 2 GB of data.

\vspace{-14pt}
\subsubsection{Synchronization:}

Apart from physical clock, each node $n$ maintains a logical clock for itself and share it with all the friends and storekeepers. Every friend and storekeeper maintains a separate clock for every user with which it's associated. Also each group owner maintains a logical clock for the community. Every community member maintains a separate clock for every community of which he is of member. Logical clock is used for keeping a synch between a node $n$ and his friends and storekeepers. On the same line it is used for keeping a synch between all the community members. It is also used to resolve multiple writes and updates (explained more in section \ref{sec:actions} either in community or for friend's comment) for a node $n$.

\vspace{-14pt}
\subsubsection{Services/Information:}
There are two kinds of services, namely - public and protected, available in the network. Anyone can avail a public service.
The list of super-peers (and the clause of their agreement) is an example of public service. In contrast, if only group members are allowed to access the information related with a particular group or community, then it will be termed as a protected service.

\vspace{-14pt}
\subsubsection{List of Directories:}
Following are the list of directories being maintained in the DOSN system.
\begin{itemize}
\item	User List : Every user's id and his public information pointers (the nodes which are storing the public data). This list is maintained by super-peers.
\item	Super-peer List: List of super-peers and their respective services and agreement details. This list is maintained by super-peers themselves.
\item	Group (or community) List: List of communities and pointers for community owner and public content of the community. This list is jointly maintained by community owners and super-peers.
\end{itemize}

\vspace{-14pt}
\subsection{Bootstrap of new node}\label{sec:protocol}
\vspace{-6pt}
A new node $n_{nn}$ joining DOSN, initially relies on super-peers to increase its data availability. Later it may find friends (or strangers) in the DOSN with help of super-peers which can act as storekeepers for it. We explain and classify different phases of a new node $n_{nn}$ from the point it enters the network till the point it settles down in the network. 

\noindent \textbf{1. Initial Phase: } \label{sec:bootstrap4node}
When a new user joins the DOSN, it creates a profile. We call this initial period as `Initial Phase' (IP). Time period of this phase could be from few hours to days. We assume (although not necessary) that the user will set his username, location and interest as public data, as these are the most used attributes to find a user in the network. A public data is a piece of information through which a user can advertise about himself to make new acquaintances. As soon as a profile is created by the user, an entry is added to userlist, which is maintained by super-peers (we note that systems like Skype follow similar approach for maintaining a distributed directory of users).

Since a new node might not get enough friends or acquaintances immediately, so it might contact some super-peers for storing his data. When a new node joins the network, he can view a list of super-peers (which is public information) and the list of services they are providing. A super-peer advertises following list of services for the new nodes.
\begin{itemize}
\item Storage: How much data they can store. In lieu of offering space to new node, in future the super-peer might take the new node's space for storing its personal data or storing some other nodes' data.
\item Agreement time: For how long they can store (e.g., for one month).
\item Time availability: What times of the day, they can cover for a node.
\item Content type: Ready to store what type of content. E.g., a super-peer might not be interested in storing any copyright material or pornographic material.
\item Advertisement: What kind of advertisements will be shown on user's profile by the super-peer.
\end{itemize}

A new node selects a super-peer and sends a request for storekeeping. Once an agreement is reached between a super-peer and the new node,
data is replicated from the new node to the super-peer. New node $n_{nn}$ now enters Take Care Phase (TCP) where a super-peer takes care of $n_{nn}$'s data.
The data at the super-peer (and at storekeepers) is stored in encrypted form if and as necessary \cite{plutus,hungACSAC} -
thus storing parties (super-peers/storekeepers) can't read or modify the content of data. 

There are number of incentives for a node to become a super-peer and thus facilitate new nodes joining the system. Super-peers get more insight in the network. It could be a self satisfaction for others, as a contributor in making the DOSN successful. Advertisements while serving other nodes' content provide opportunities for monetization. Users in turn give feedback about super-peers at the end of contract which can affect a super-peer's reputation, thus reputation is another source of incentive for super-peers. A higher reputation will in turn enable super-peers to attract more users, leading to more influence and further monetization opportunities for the super-peer.

\noindent \textbf{2. Take Care Phase:}
Once an agreement is reached between $n_{nn}$ and a super-peer, $n_{nn}$ enters Take Care Phase (TCP), in which the super-peer helps improve $n_{nn}$'s data availability and thus acts as a storekeeper. Note that, the super-peer may delegate the store-keeping task to other nodes who owe the super-peer because the super-peer had previously helped them when they were new in the system.

The take care period is the time when the new node tries to find and establish friendships in the system. There are various ways through which a node can add new friends. Node $n_{nn}$ can try to search for his acquaintances or friends from userlist in the system and also through out-of-channel mechanisms such as emails to know if they are participating in the network. By exploring community's membership, a user can also make new friends based on shared interests. A node tries to find suitable storekeepers among all the friends\footnote{One compelling factor for choosing storekeepers could be their uptime, however in general this is not always possible. As in real life, we can't compel our acquaintances/friends to store data.}.

As part of agreement, super-peers keep track of new node's geographic location and up-time statistics. Each super-peer maintains two kinds of pools as describe below with the objective to facilitate users in finding other nodes, which are complete strangers to them and can act as storekeepers and thus help in increasing data availability.
\begin{itemize}
\item Time Track pool (TT): As part of agreement, during take care phase, the super-peer tracks new node's up and down time.
\item Stranger pool: Every super-peer maintains a stranger pool along with TT pool. A node can make an explicit request to a super-peer to ask for suggestions about the nodes which can act as storekeepers for it. During such a request to the super-peer, a node submits it's own up and down time and specifies the time intervals during which it wants the store-keepers to cover for it.
\end{itemize}

All super-peers share TT and stranger pool information among themselves. If a node doesn't get enough uptime availability from its friends, it can request super-peers to make some suggestion for it by using above two pools. Based on the two pools, a super-peer can suggest requesting nodes about the (stranger) nodes which can probably act as storekeepers\footnote{Nodes suggested by super-peer based on both the pools are termed as stranger nodes.}. Once the suggestions are received about other nodes, a node can request these suggested nodes for storing data. Once an agreement is reached between two nodes, data is replicated and stored securely.

\noindent \textbf{3. Settled Phase: }
A node may extend its take care phase with a super-peer if he is not able to get enough storekeepers of its own to get desired data visibility. Once a node has enough storekeeper(s) (other than super-peer) to manage his data, it can relinquish super-peer's services. For every node $n$, every storekeeper agrees to store a limited amount of data which requires space $s$. So only latest $s$ amount of data is stored with storekeepers and once the limit is reached, the old data is deleted from storekeepers.

\vspace{-12pt}
\subsection{Data Synchronization and Updates.} \label{sec:actions}
\vspace{-9pt}
This section describe merging, copying and updates of data between storekeeper(s) which is one of the important aspects of DOSN.

\vspace{-15pt}
\subsubsection{Data Storage: }
When a node $n$ goes down it pushes all its latest data to all the up storekeepers. When node $n$ comes up, it requests all the storekeepers (whoever are up) for the latest data. It accepts the data from the storekeeper, which is having latest data. Node $n$ could be acting as storekeeper for other nodes as well. As a storekeeper, when a node comes up it synchronizes the data (for each of node, for which it is acting as storekeeper) with other storekeepers or with owner of the data (whoever is up) for which it is acting as storekeeper.

If two storekeepers don't have enough intersecting up time, then a node can request a super-peer to act as temporary data storage for the latest data. Thus in this case a super-peer acts as a rendezvous point and keeps the latest data from the last up storekeeper which is going down before another storekeeper comes up (with respect to node $n$). We assume that, before logging off, a storekeeper do transfer completely the latest data on another machine (either super-peer or storekeeper). In this case, super-peer doesn't take part in any updates or cater any request from friends accessing the data. When node $n$ comes up and if no storekeeper is up, then node $n$ requests the designated super-peer for the latest data.

When node $n$ is down, friends of $n$ contact respective storekeepers to access the latest data for node $n$. Any of the storekeepers who are up can be contacted with a request for the latest data. A particular storekeeper is chosen randomly from up storekeepers for data access.

\vspace{-12pt}
\subsubsection{Updates: }
When a friend \emph{f}, posts a comment on node $n$'s status (or about new uploaded picture or link), an update request is send to node $n$ (if its up) with a timestamp (logical timestamp). If $n$ is down, requests are made to all the storekeepers who are up, about the updates. Every update request comes with the timestamp \footnote{Timestamp is the time of the logical clock w.r.t. node $n$ which is shared by friends and storekeepers of node $n$. Since the updates are w.r.t. node $n$, so logical clock of node $n$ is used.}. A node $n$ (or storekeeper(s)) might receive multiple updates at same item, in such cases, node $n$ (or storekeeper(s)) uses timestamp to resolve any conflicts.

If neither node $n$ is up and nor any of its storekeeper, then comments/posts on the wall are stored on the commenter side, and resides till node $n$ (or any of its storekeeper) comes up.

\vspace{-12pt}
\subsection{System Model}\label{sec:system-model}
\vspace{-6pt}
Let $G(N,E)$ be the DOSN graph, where $N$ represents nodes (or users) of the network and $E$ represents the set of edges between any two nodes $n_{i}$ and $n_{j}$ $\in$ $N$. An edge e $\in$ $E$ exists between $n_{i}$ and $n_{j}$ $\in$ $N$ if each $n_{i}$ and $n_{j}$ profile's are linked with each other as friends.

Let $\Psi$ represent the global stranger pool managed by all super-peers and $SP$ the list of super-peers. To increase the availability, a new node can share his data with a particular super-peer $n_{sp}$ $s.t$ $n_{sp} \in SP$. A node can get good availability with the help of friends if many of them are ready to act as storekeepers. We define friends of node $n_i$ as the set $\gamma_i$ where $n_f \in \gamma_i$ \emph{iff} $e_{if} \in E$. A node might be able to convince some strangers for storekeeping to boast his availability. Let $\sigma_i$ represent the set of storekeepers for node $n_{i}$, then formally it can be represented as
\vspace{-5pt}
\begin{center}
            $\sigma_i = n_{i} + \sum_{sk=1}^{SK} n_{sk}$
            \\
            $s.t$ $ \forall n_{sk}$ $n_{sk} \in \gamma_i$ or $n_{sk} \in \psi$ or $n_{sk} \in SP$
\end{center}
\vspace{-5pt}
Let $T_{i}$ represents the set of time intervals for which node $n_{i}$ is online. In decentralized system if the data is not stored at other than the owner of the data, then data availability is equal to the time units for which node $n_{i}$ is up, that is Avl($n_{i}$) = $|T_{i}|$. If a node is able to store data on other nodes as well, then availability of node $n_{i}$ can be defined as
\vspace{-5pt}
\begin{center}
            $Avl(n_i)$ =  $|\bigcup_{al=1}^{AN}T_{al}|$ $s.t$ $\forall n_{al} \in \sigma_i$
\end{center}
\vspace{-5pt}

\vspace{-19pt}
\section{Evaluation}\label{sec:eval}
\vspace{-15pt}
In this section we evaluate using various metrics how a DOSN based on SuperNova architecture will perform for different mix of user behaviors. We first provide a description of various parameters for user's behavior, before discussing the results.

\vspace{-18pt}
\subsection{Network }
\vspace{-14pt}
We use the social network graph from a subset of the DBLP co-authorship graph as a representative social network. Specifically, we use the giant connected component of co-authorship graph from DBLP record of papers published in conferences between 2004 to 2008, comprising of 273798 unique authors. Each author represents a node in the network. The graph has an average degree of 6 with diameter of 23.

\vspace{-12pt}
\subsection{Trace Generation }
\vspace{-7pt}
Node's up and down time is generated synthetically, taking into account typical bimodal, and weekly behavior cycles. A trace was generated for two weeks time. We considered following three different factors for the trace generation :
\begin{itemize}
\item Location: We consider twelve different (time zone) locations, with each location differing by 2 hours from its adjacent locations.
\item Weekday Behavior: We consider six types of behaviors during the weekday.
\item Weekend Behavior: We consider four types of behaviors for the weekend.
\end{itemize}

Weekday's and Weekend's behavior were motivated by different working hours which users might follow on their workstations and thus will drive the online time during weekdays. Also some users work during weekends and others simply leisure and don't switch on their workstations. Each user is associated with one of each three different factors and a resultant trace was attached with that user for one week. In the second week, we consider two scenarios: (i) all users have same behavior as in the first week, and (ii) for some users (27 percent) we introduce minor variations in their traces and for few others we generate a major variations (3 percent), while for the rest (70 percent) the behavior remains unchanged.

\vspace{-12pt}
\subsection{Various parameters for Storekeepers }
\vspace{-7pt}
In our study we consider three different parameters related with acceptance of node's request for storekeeping the data. Different users have different behavior and to capture individual behavior is difficult so we generalize it for the whole network.
\begin{itemize}
\item $P_{fd}$: This parameter decides with what probability a friend/neighbor is ready to accept the data of a node.
\item $P_{sd}$: The percentage of nodes which wants to participate in stranger pool.
\item $P_{as}$: With what percentage a stranger is ready to accept another stranger's request for storing the data.
\end{itemize}

We consider two values for each parameter - representing low and high percentage values taken by the nodes for particular storekeeping behavior. Based on these values we consider 8 different combinations. Table \ref{Tbl:ParamComb} summarizes the various combinations that we analyzed by selecting different percentage values for each parameter. The first column represents the combination number to identify the parameter combinations, and $2^{nd}$, $3^{rd}$ and $4^{th}$ columns represent the percentage values of each parameter.

\vspace{-12pt}
\subsection{Various node strategies for storage }
\vspace{-7pt}
A node can request social neighbors or/and strangers for storekeeping. It is less likely that a new joinee will make enough friends/strangers to store its data during few initial days, thus new nodes are more likely to request super-peers for data replication and thus will choose one of the super-peers for sharing of data\footnote{For simplicity in our experiments we assume a node only requests one super-peer for storing its data.}. Nodes will try their best to have maximum availability by replicating their data on various nodes across the network. A node in general can store its data using either of the four following schemes depending on various factors like if its social neighbors are already overloaded and whether he is (not) willing to store data on nodes run by strangers.
\vspace{-5pt}
\begin{itemize}
\item $S_{nns}$: A node is able to store his data on some of his neighbors as well on strangers.
\item $S_{nn}$: A node is able to store data only at his neighbors. A node might not be interested in storing data at strangers because either he is not trusting or he is not able to convince strangers.
\item $S_{nsp}$: A new nodes store his data on one of the super-peers after agreement.
\item $S_{ns}$: A node might not be able to get help from any of its neighbors for storage, e.g., because all the neighbors are already overloaded because of prior storage commitments. However a node might go for a scheme where he is able to convince some strangers for storing the data.
\end{itemize}

Super-peers are assumed to have enough storage space to cater to node's request (specially the new joinees). By varying $P_{fd}$, $P_{sd}$  and $P_{as}$, we measured the percentage of nodes falling for $S_{nns}$, $S_{nn}$, $S_{nsp}$ and $S_{ns}$. In Table \ref{Tbl:ParamComb} the $5^{th}$, $6^{th}$, $7^{th}$ and $8^{th}$ columns represent the percentage values for $S_{nns}$, $S_{nn}$, $S_{ns}$  and $S_{nsp}$.

\vspace{-10pt}
\begin{table}
  \caption{Different combination of Parameters}
  \vspace{-10pt}
  \begin{center}
  \begin{tabular}{| c | c | c | c | c | c | c | c | }
    \hline
      Combination &  $P_{fd}$ & $P_{sd}$ & $P_{as}$ &  $S_{nns}$ & $S_{nn}$ & $S_{ns}$ & $S_{nsp}$  \\ \hline
       C1        &  50 & 50 & 100                   &  22.4      & 36.6     & 16.2     & 24.8       \\ \hline
       C2        &  50 & 70 & 100                   &  31.5      & 27.7     & 22.7     & 18.1       \\ \hline
       C3        &  20 & 70 & 100                   &  14.5      & 11.9     & 41       & 32.6       \\ \hline
       C4        &  20 & 50 & 100                   &  10        & 15.9     & 29.2     & 44.9       \\ \hline
       C5        &  20 & 50 & 40                    &  4         & 22       & 12       & 62         \\ \hline
       C6        &  20 & 70 & 40                    &  5         & 20       & 16       & 59         \\ \hline
       C7        &  50 & 70 & 40                    &  12        & 46       & 9        & 33         \\ \hline
       C8        &  50 & 50 & 40                    &  9         & 45       & 6        & 40         \\ \hline
  \end{tabular}
  \end{center}
  \label{Tbl:ParamComb}
  \vspace{-15pt}
  \end{table}

\begin{figure*}[t]
\center{\includegraphics[width=0.9\textwidth]{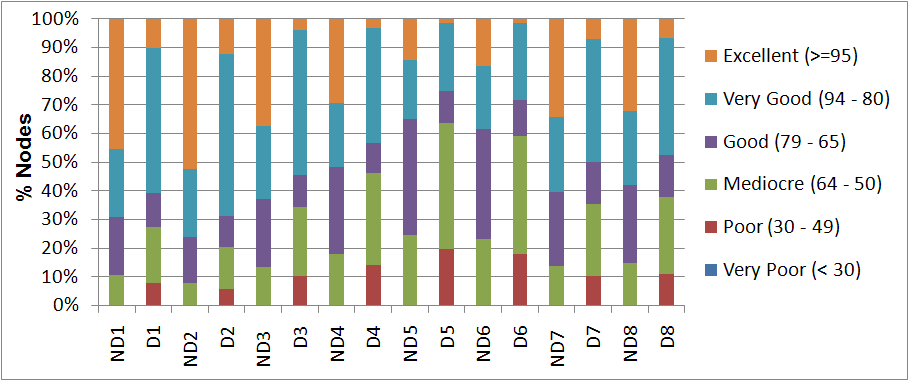}}
\vspace{-6pt}
\caption{Performance of different parameter combinations for Deviation (D) and Non Deviation (ND)}
\vspace{-12pt}
\label{fig:all-comb-perf}
\vspace{-9pt}
\end{figure*}

We also measure the availability of nodes in each combination. We define availability time as the ratio of the number of time units for which a node $n$'s data is available to that of total time unit of the simulation. We categories the availability (Avl) into 6 classes namely (names are intuitive for performance) - (1) Excellent $(Avl \geq 95)$ (2) Very Good $(95 > Avl \geq 80)$ (3) Good $(80 > Avl \geq 65)$ (4) Mediocre $(65 > Avl \geq 50)$  (5) Poor $(50 > Avl \geq 30)$ and (6) Very Bad $(Avl < 30)$. For each combination we measure the percentage of nodes falling under each availability category. Figure \ref{fig:all-comb-perf} represents the percentage values for each combination of parameters namely from C1 to C8 (refer Table \ref{Tbl:ParamComb}) for both cases when there is deviation (D) and when there is no deviation (ND).

Parameter combinations representing the user's behavior has clear effect on the percentage of nodes falling under each scheme. In an ideal situation (Row C2 in Table \ref{Tbl:ParamComb}- based on our parameters), half of neighbors (or friends) ( $P_{fd}$ = 50\%) are ready to accept storage request and more than half of the nodes in the network ($P_{sd}$ = 70\%) are ready to store participate in stranger pool and all the strangers ($P_{as}$ = 100\%) are ready to accept all the requesting for storage. In this situation there is minimum load on super-peers (comparing with all other cases). Also the number of nodes falling under excellent availability category is at its maximum. Although deviation has introduced a sharp decline in percentage of excellent nodes however the total numbers of excellent and very good nodes are still higher. When the percentage of nodes ready to store data at strangers declines ($P_{sd}$ in C1) there is a very small effect on the excellent nodes (compare ND1 and ND2 in Figure \ref{fig:all-comb-perf}).

However when the participation of number of nodes willing to participate in stranger list declines (from 70 \% to 50 \% for $P_{sd}$ in Row C1) we notice an immediate load on super-peers (Row C1, Column $S_{nsp}$). This is better than C7, which represent the situation when the percentage of nodes willing to store stranger's data decreases. Thus presenting the fact that parameter $P_{as}$ is more influential than $P_{sd}$ for increasing the load on super-peers.

Maximum load (62\%) on super-peers (Row C5, Column $S_{nsp}$) is observed when neither neighbors are interested in storing other friend's data and nor the participation of nodes in stranger's scheme is overwhelming (Row C5). Although storage is generally cheap, in practice, super-peers nevertheless will have limited space. Furthermore, there is also the issue of associated bandwidth usage. Hence, such a system state is undesirable and arguably unsustainable.

There should be a balanced behavior among nodes for a system to be successful. When the percentage of neighbors, who are willing to store neighbor's data is low (Row C3 and C4), the system can still sustain if there is high level of contribution by/to strangers. This depicts a scenario if all new joinees act as storekeepers for each other, thus increasing availability. This collaboration of stranger support (in time of need) may eventually lead to change in relationship from strangers to friends.

We noticed high load on super-peers when percentage of the nodes falling under stranger scheme is high and when percentage of nodes ready to accept stranger data is low (Row C6). When participation of neighbors is increased, there is immediate decline on super-peers' load as evident from load Row C7 which is attributed due to the fact that nodes rely more on neighbors than strangers. By comparing C6 and C7 we infer that merely participation of nodes in stranger scheme ($P_{sd}$) is not sufficient however it is more important that how many nodes out of $P_{sd}$ are ready to accept stranger's data ($P_{as}$), thus parameter $P_{as}$ is more important than $P_{sd}$. The system shows a very poor performance when all the parameters are at their lowest (Row C5) - clearly showing the decline in percentage of excellent and very good availability (D5 in Figure \ref{fig:all-comb-perf}).

Although nodes are more willing to store at friends, for new nodes and the nodes who are not able to find storage among neighbors and strangers, super-peer can be of great help. From above values we observed that the system has good performance when everyone cooperates (C2) and is at its worst when users are not ready to contribute to the system (C5) and act in a selfish manner.

We compare availability in three different scenarios. (1) To highlight the effect of super-peers we compare our super-peer based architecture with a flat scheme. (2) A comparison between an ideal (or best) case vs worst is also presented to reflect the effect of cooperation and selfishness respectively. (3) Nodes might be interested in getting availability in terms of the time when their friends are online vs. 24/7 availability. In evaluation of all the three cases - along with measuring cumulative availability, we try to gather information about what percentage of nodes are getting what percentage of availability. As a micro-measurement, we also compare the individual scenarios in terms of availability in terms of 6 categories defined above.

\begin{figure*}[t]
  \begin{center}
    \subfigure[Cumulative Availability]
    {\label{fig:CmlAvl-FandT}\includegraphics[width=0.49\textwidth]{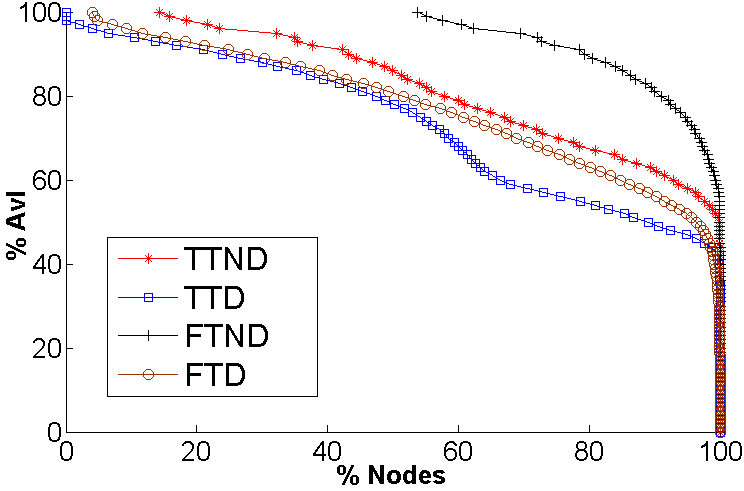}}
    \subfigure[Individual Availability]
    {\label{fig:IndAvl-FandT}\includegraphics[width=0.49\textwidth]{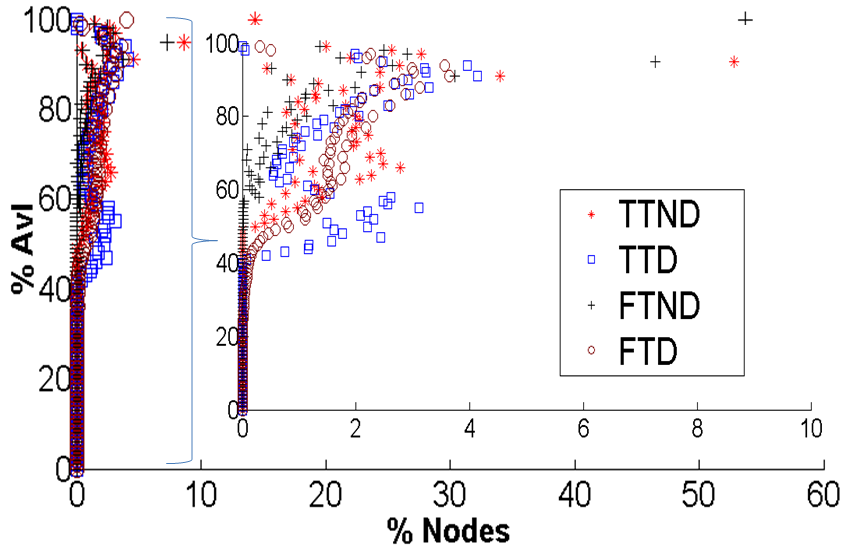}}
    \subfigure[System Performance]
    {\label{fig:CatgAvl-FandT}\includegraphics[width=0.65\textwidth]{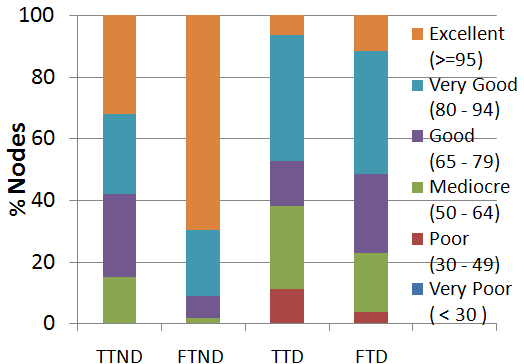}}
  \end{center}
  \vspace{-10pt}
  \caption{Comparison for Friend's Time (FT) and Total Time (TT) for Deviation (D) and NonDeviation (ND)} \label{fig:AvlFandTT}
  \vspace{-15pt}
\end{figure*}

\noindent \textbf{1. Availability w.r.t total time and time when friends are online: }
A node could be interested in making his data available only when his friends are online (so that they can access it), instead of getting 24/7 availability.We selected parameter combination in Row 8 (which is the most realistic scenario among all the cases) to compare two different types of availabilities. Interesting results were gathered while comparing two different cases.  Figure \ref{fig:AvlFandTT} compares availability for the two scenarios, i.e., when total time (TT) is considered and when friends up time (FT) is considered. Figure \ref{fig:CmlAvl-FandT} gives the cumulative availability when deviation (D) is introduced as well as when system is evaluated with no-deviation (ND). We present our results for a pragmatic scenario (C8 in Table \ref{Tbl:ParamComb}) out of all the combinations.

When a node is only concerned about availability of its data when his own friend's are online, and there is no-deviation in the node's uptime behavior (FTND), the system outperforms the total time with no-deviation (TTND). When only friend's up time is considered there are more than 50\% of nodes whose availability is 100\% as compare to total time (TT) where the value is less than 20\% (refer Figure \ref{fig:CmlAvl-FandT}. However with introduction of deviation, in case of friends up time (FTD) system sharply declines in terms of performance as compare to total time (TTD), where deviation doesn't put much effect on the performance. However in both cases all nodes had availability greater than 40\%. In Figure \ref{fig:IndAvl-FandT} we capture more microscopic information in terms of the percentage of nodes getting a specific percentage of availability. This information is important to gather individual behavior as some information is missing in the macroscopic cumulative distribution. In case of FTND the highest percentage of nodes is around 53 \% however it declines to around 5\% when deviation is introduced. From the Figure \ref{fig:IndAvl-FandT}, one can also infer that there is not much deviation in terms of individual percentage after deviation is introduced.

It's common in real world that most friends are collocated. So if a node stores the data among his friends, because of location (and consequent temporal) correlation, most of the friends will be online during the same time and thus availability will be high for such nodes. For same reason, most of the nodes get excellent availability. However the value might be low if we compare the total time of the system which is much higher than the friend's online time and thus a decrease in availability value is observed. When deviation is introduced, there is a sharp decrease in availability. If friends deviate from their usual uptime behavior they are responsible for decreasing the data-availability as compared to a node which has scattered his data, like by selecting some strangers as well. This observation is reinforced with the help of Figure \ref{fig:CatgAvl-FandT}, where the total percentage of nodes for "excellent" plus "very good" in case of friends after introducing deviation (FD) becomes less than the total excellent nodes without introducing the deviation (FND). If we consider the total time as a measure to find availability, we notice that there is not much difference in performance (excellent plus very good) between no deviation (NDTT) and when deviation is introduced (TTD). This can be attributed to the fact that if a node spread his data to nodes other than friends, then he will be less affected by the deviation and thus he can either choose strangers, and if strangers are not willing to do so he can take help of super-peers.

\noindent \textbf{2. Comparison of Flat Vs Super-peer architectures:}
We also compare our approach to a flat scheme where there are no super-peers, in contrast to our architecture. In the absence of a trusting authority like super-peers, it is difficult to keep a strangers' list or trust information regarding strangers for finding suitable store-keepers. However, to do a fair comparison with the flat scheme, we assume that nodes can convince strangers to store their data using a reciprocity scheme \cite{krzicdcs10}. This assumption works well with the nodes which are present in the system for long. However, the new joinees find it difficult without a trust authority (like super-peers) to decide which strangers to trust. This affects the overall performance of the system as the new joinees are not able to get enough storekeepers for data-replication.

\begin{figure*}[t]
  \begin{center}
    \subfigure[Cumulative Availability]
    {\label{fig:CmlAvl-FandSP}\includegraphics[width=0.49\textwidth]{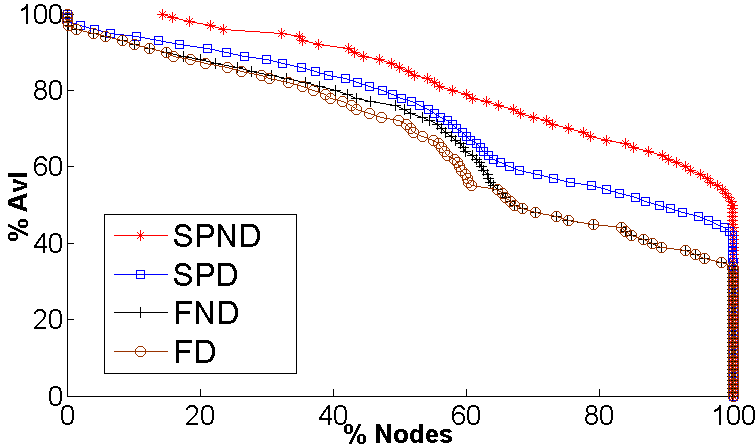}}
    \subfigure[Individual Availability]
    {\label{fig:IndAvl-FandSP}\includegraphics[width=0.49\textwidth]{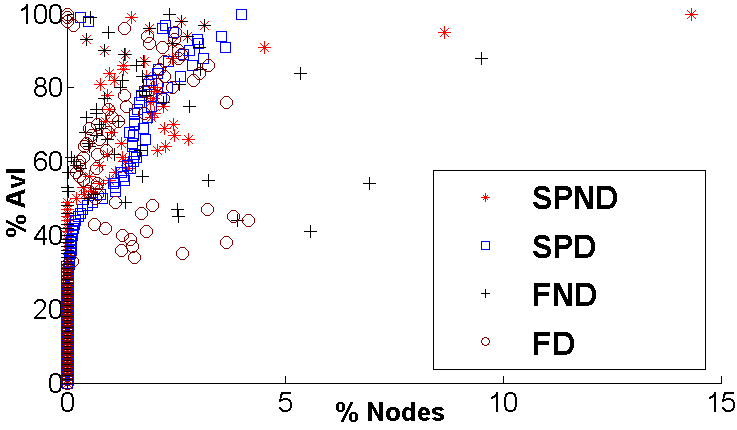}}
    \subfigure[System Performance]
    {\label{fig:CatgAvl-FandSP}\includegraphics[width=0.65\textwidth]{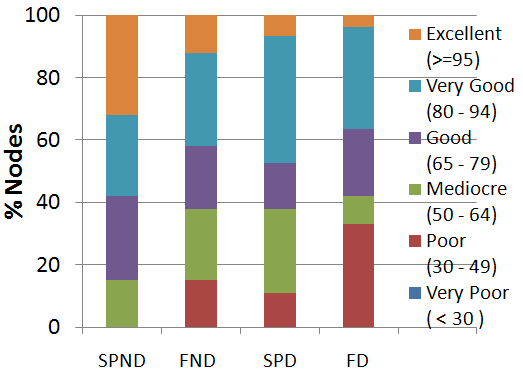}}
  \end{center}
  \vspace{-10pt}
  \caption{Comparison for Flat (F) Vs Super-peer (SP) for Deviation (D) and NonDeviation (ND)} \label{fig:AvlFandSP}
  \vspace{-15pt}
\end{figure*}

As in the bootstrap period, it is difficult for a new node to find friends and convince strangers to store their data and this is a major cause of poor performance in the flat scheme. Figure \ref{fig:CmlAvl-FandSP} and \ref{fig:IndAvl-FandSP} show that there are very few nodes whose availabilities are above 60 \%. Deviation (FD) downgrades availabilities for 30 \% of nodes (refer Figure \ref{fig:CmlAvl-FandSP} - between 30\% to 60 \% nodes on X axis). We further observe that only a small percentage of nodes (~8 \%) have excellent availabilities, and it becomes less than half in the case of deviation (FD). These results indicate the efficacy and need of our super-peer based architecture.

\noindent \textbf{3. Best Vs Worst case: }
We also compare the best parameter (B) combination with the worst (W) one (C2 and C5 respectively in Table \ref{Tbl:ParamComb}) out of our choice of parameters. In the worst case, the low parameter values indicate that every peer is acting in a selfish and conservative manner. Evidently, super-peers are having maximum load in this scenario out of all (refer to Table \ref{Tbl:ParamComb}, Row C5 and column $S_{nsp}$).

\begin{figure*}[t]
  \begin{center}
     \subfigure[Cumulative Availability]
    {\label{fig:CmlAvl-BandW}\includegraphics[width=0.49\textwidth]{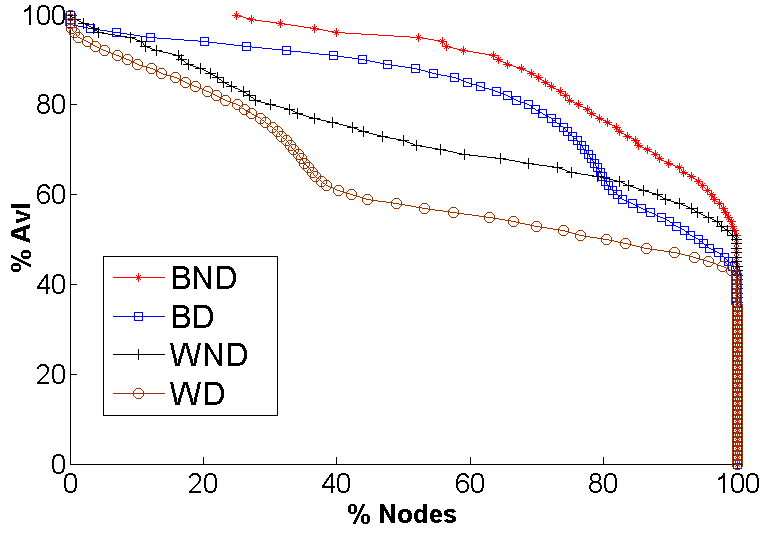}}
    \subfigure[Individual Availability]
    {\label{fig:IndAvl-BandW}\includegraphics[width=0.49\textwidth]{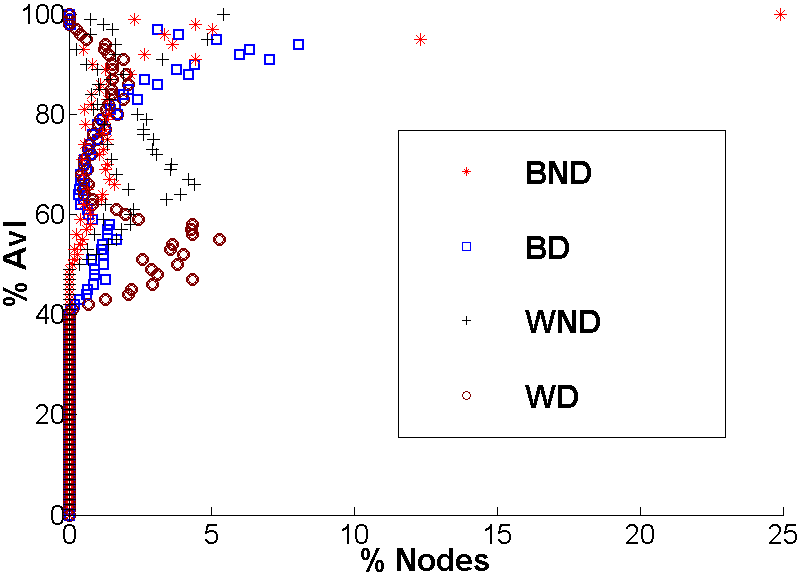}}
    \subfigure[System Performance]
    {\label{fig:CatgAvl-BandW}\includegraphics[width=0.65\textwidth]{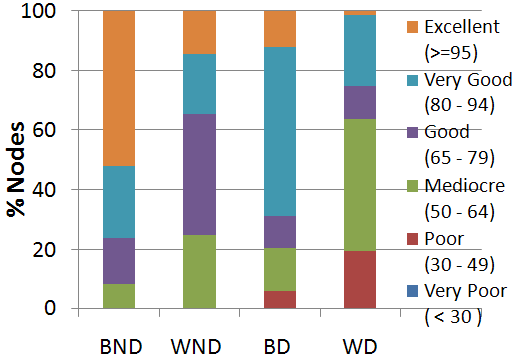}}
  \end{center}
   \vspace{-10pt}
  \caption{Comparison for Best(B) Vs Worst(W) Parameter Combination for Deviation (D) and NonDeviation(ND)} \label{fig:AvlBandW}
  \vspace{-15pt}
\end{figure*}

In our experiments, while super-peers have uptime comparatively higher than normal peers, they are still not individually up 24/7. Since large numbers of nodes are relying on few nodes (super-peers) when the corresponding super-peers are down, the data availability for large number of nodes decreases, thus lowering the performance of the system. Figure \ref{fig:CmlAvl-BandW} measures the cumulative values in worst case for both deviation (WD) and non-deviation (WND) and compare them with the best case for both deviation (BD) and non-deviation (BND). These values were compared w.r.to total time. Note also that the bandwidth load at super-peers is expected to increase proportionally to their storage node, which may in turn lead to congestion and further deterioration of response times and quality of service.

The low parameter values in Row C2 (for $P_{fd}$, $P_{sd}$ and $P_{as}$) shows that nodes are unwilling to act as storekeepers for others. From Figure \ref{fig:IndAvl-BandW} one can clearly note that the highest availability is around 90\% for little more than 5 \% of nodes as compared to the nearly 25 \% when every node cooperates in case of non-deviation (in best case).
The cooperation phenomena is also evident in terms of percentage of nodes getting "excellent" or "very good" availability both with and without deviation for the best case in Figure \ref{fig:CatgAvl-BandW} (BD and BND respectively). In the worst case, after introducing deviation the system is behaving more in a mediocre range where storekeepers might not be able to exchange latest data thus leading to inconsistency. Also from Figure \ref{fig:CmlAvl-BandW} it's easy to infer that the availability is 60 \% or less for 60 \% of the nodes. Even without deviation, most of the nodes are not getting good availability as around 50 \% nodes are having availability 70 \% or less.

\section{Conclusion and Future Work}\label{sec:concl}
\vspace{-10pt}
In this paper we proposed a super-peers based architecture to facilitate decentralized online social networking. The proposed architecture provides flexibility to individual users to choose their strategy in terms of where to store their data, as well as whose data to store. While we do not investigate the role of trust and reputation explicitly in this paper, recent trust model \cite{superpeertrust} which correlate node's high reputation with them being more influential custodians of the community's tasks provides a premise to carry out such a dynamic analysis of node's behavior in a SuperNova like super-peer based DOSN. In this paper, we instead investigate the `static' characteristics of the system, specifically, we study the influence of different user strategy mixes on the performance and sustainability of a DOSN. Existing initiatives, both academic and open-source ones, have in recent years focused on developing the enabling technologies (some of which, particularly related to security, could be incorporated in SuperNova as well), however, a quantitative estimate on whether such a system is sustainable or not, given diverse participation behaviors and capacities has so far not been investigated. Thus, besides proposing a more pragmatic system design, this paper distinguishes itself from existing work in the rigor of the evaluations. Studying the dynamic role of reputation in changing population mixes, as well as designing and incorporating incentive mechanisms in SuperNova in order to achieve desirable mix of user populations; refinement of the security mechanisms in the design; and incorporation of the SuperNova mechanisms in the PeerSoN initiative comprise our planned future work.

\section*{Acknowledgements}
\vspace{-10pt}
The work in this paper has been funded in part by NTU/MoE's AcRF Tier-1 RG 29/09. We thank Deepak Subramanian, Tu Trung Hieu and Anthony Ventresque for their comments and feedback in general on this work.

\bibliographystyle{abbrv}
\bibliography{SuperNova}
\label{sec:References}

%
\end{document}